\documentclass[10pt,onecolumn]{article}
\usepackage{amsmath}
\usepackage{graphics}
\usepackage{graphicx}
\usepackage{amsfonts}
\usepackage{amssymb}

\begin{document}

\title{Substituting a qubit for an arbitrarily large number of classical bits}
\author{Ernesto F. Galv\~{a}o and Lucien Hardy\\\textit{Perimeter Institute for Theoretical Physics,}\\\textit{35 King Street North, Waterloo, Ontario, N2J 2W9 Canada.}}
\maketitle

\begin{abstract}
We show that a qubit can be used to substitute for an arbitrarily
large number of classical bits. We consider a physical system
$S$ interacting locally with a classical field $\varphi(x)$ as it
travels directly from point $A$ to point $B$. Our task is to use $S$
to answer a simple yes/no question about $\varphi(x)$. If $S$ is a
qubit, the task can be done perfectly. On the other hand, if $S$ is a
classical system then we show that it must carry an arbitrarily large
amount of classical information. We identify the physical reason for
such a huge quantum advantage, and show that it also implies a large
difference between the size of quantum and classical memories
necessary for some computations. We also present a simple proof that
no finite amount of one-way classical communication can perfectly
simulate the effect of quantum entanglement.
\end{abstract}


\maketitle

A general pure state of a two-level quantum system -- a qubit -- can
be written as
\begin{equation}
a|0\rangle+b|1\rangle, \label{qubit state}
\end{equation}
where $a$ and $b$ are complex numbers satisfying
$|a|^{2}+|b|^{2}=1$. In general, in order to specify these continuous parameters to arbitrary accuracy  we need an arbitrarily large number of classical bits of information. This unbounded amount of information contained in a single state would seem to suggest that a qubit could be used to communicate an arbitrarily large amount of information. Of course, this turns out not to be possible: Holevo's bound \cite{Holevo73} implies that a single qubit cannot be used to transmit more than one bit of information  between two previously unentangled parties.

This raises doubts about the objective existence of the information encoded in the continuous parameters in (\ref{qubit state}). In this paper we show how this large amount of information can be used in an information processing task. The trick
is not to require that this information can actually be read out, but
rather to use it to substitute for a large amount of classical
information. We illustrate this with a very simple information
processing task which can be performed perfectly with a single qubit
but would require an arbitrarily large amount of classical
communication. From a computational perspective, we will show that
this means that for some tasks quantum computers can offer an
unbounded advantage in memory size, while still being as
time-efficient as classical computers.

The idea of using quantum communication to substitute for classical
communication was pioneered by Yao \cite{Yao93} and further developed
by various authors \cite{BuhrmanCW98, ClevevDNT99, Raz99}. Their
work established that quantum communication can be substantially
better than classical communication (see \cite{Brassard01} for a survey). In addition, in \cite{CleveB97}
Cleve and Buhrman showed that quantum entanglement could substitute
for classical communication, an application which was further explored
in other papers \cite{Grover97, BuhrmanCvD01, BuhrmanvDHT99}. In
relation to this quantum resource, we will show that no finite amount
of one-way classical communication can perfectly simulate the effect of entanglement.

\noindent\textbf{The task. }We now come to the information processing
task. We have a one dimensional real field $\varphi(x)$ defined on the
straight line between points $A$ and $B$ in the $x$-axis. The integrated value of the
field is guaranteed to be equal to an integer, $m$, times a real constant, $\alpha$:
\begin{equation}
\int_{A}^{B}\varphi(x)dx=m\alpha. \label{constraint}
\end{equation}
The task is simply to find out whether $m$ is odd or even by sending
some physical system $S$ directly from point $A$ to point $B$ (it is not
allowed to move backwards). We can choose any local coupling of the
system with the field.

Next we will present simple quantum and classical
protocols for solving this problem. The particular classical protocol
we present uses a system $S$ with an infinite number of distinguishable states, i.e. corresponds to communication of an arbitrarily large number of bits. The quantum protocol, on the other hand, requires only a single qubit. After the two examples of protocols, we will show rigorously
that this huge difference in communication power holds between the
quantum protocol and \textit{any} classical protocol.

\noindent\textbf{The quantum protocol.} Let the quantum system $S$ be
a spin-half particle initially prepared in the spin up (along the $z$
direction) state. Now allow the field to rotate the spin in the $y-z$
plane by an amount proportional to the strength of the field. This
constant of proportionality can be chosen to be such that, by the time
$S$ reaches $B$, the spin has been rotated through $m/2$ full
rotations. Thus, the spin will be up if $m$ is even and down if $m$ is
odd. By measuring the spin along the $z$ direction at $B$ we can
distinguish these two possibilities and solve the task.

A simple realization of this protocol using the polarization of a photon
is the following. The classical field $\varphi(x)$ can be taken to be
the magnetic field $B_{x}=\overrightarrow{B}\cdot\hat{x}$ produced by
a non-uniform current density in a solenoid wrapped around a
transparent rod. The field $B_{x}$ is proportional to the local
current density. If a photon with vertical polarization is sent
through the rod, its polarization vector at the other end will have
been rotated by an amount proportional to the integrated value of the
field (due to the Faraday effect). The constant of proportionality
associated with the interaction can be altered by changing the
physical properties of the transparent rod. The polarization can be
analysed at $B$ along the vertical-horizontal directions. If the
constant of proportionality is correctly adjusted, the polarization at
$B$ will be vertical if $m$ is even and horizontal if $m$ is odd,
solving the task.

\noindent\textbf{A classical protocol.} To solve the problem with a
classical system, let $S$ be a rod pivoted at one end and free to
rotate in the $y-z$ plane. The position of the rod is given by a real
angle $\theta$ whose specification requires and infinite number of
bits. In other words, the rod is a classical system with an infinite
number of distinguishable states.

We start with the rod in the up direction. The rod-field coupling is chosen so that as the rod moves through distance $dx$ it is rotated by $\eta\varphi(x)dx$. For a coupling constant of $\eta=1/(2\alpha)$, the rod will be rotated by exactly $m/2$ turns at the end. This means that it will be pointing up if $m$ is even and down if $m$ is odd. Even in the presence of some noise or jitter, we can still distinguish these two possibilities as long as the accumulated error is not greater than plus or minus a quarter of a rotation. This tolerance to noise indicates that a more careful analysis is required before we conclude that we need a classical system $S$ with an infinite number of distinguishable states, such as the rod.

\noindent\textbf{The general classical case.} Let us now prove that the stated task cannot be solved using any classical system $S$ possessing only a finite number of distinguishable states.

We start by dividing the interval $AB$ into $N$ equal sections, each
of length $1/N$, and label them sequentially from $1$ to $N$, starting
from $A$. Now define $\phi_{n}$ to be the integrated value of the
field $\varphi(x)$ over the $n$th section. Constraint (\ref{constraint}) now reads:
\begin{equation}
\sum_{n=1}^{N}\phi_{n}=m\alpha. \label{summalpha}
\end{equation}

We do not need to impose any restrictions on the nature of the
function $\varphi(x)$ except that it can be integrated. If we want
$\varphi(x)$ to correspond to a physical field it must be continuous,
but that is not required for our proof to work.

Even with continuous $\varphi(x)$, there is no constraint on the numbers
$\phi_{n}$ except that they sum to $m\alpha$. Hence, we can find
fields $\varphi(x)$ which correspond to any point in the coordinate
space $\{\phi_{n}\}$ for $n=1$ to $N-1$, with $\phi_{N}$ being chosen
so that eq. (\ref{summalpha}) holds.

If this communication task can be
solved for an arbitrary field $\varphi(x)$, then it can be solved for
any subset of points in the coordinate space $\{\phi_{n}\}$. Hence, we can
further simplify the problem by allowing the $\phi_{n}$ to take only a discrete set of values ${\alpha k_{n}/K}$, where $k_{n}\in\{0,1,...,2K-1\}$ and $K$ is a power of two, for a technical reason that will become apparent later. Note that $\phi_{n}$ takes values in the range $0$ to $\alpha(2-1/K)$. This makes sense as the sum is only important modulo $2\alpha$ (since we are only interested in whether $m$ is odd or even).

From now on we will refer to the $k_{n}$'s instead to the $\phi_{n}$'s.
Condition (\ref{summalpha}) becomes
\begin{equation}
\sum_{n=1}^{N}k_{n}=mK.
\end{equation}
Again, since we are only interested in whether $m$ is odd or even we
will only be concerned with sums over the $k_{n}$'s modulo $2K$.

Let us now establish a notation that enables us to discuss all
possible classical protocols involving communication with a
finite-dimensional classical system $S$. Let $L$ denote the number of
distinguishable states of $S$. Communication with a classical system
with $L$ distinguishable states is the equivalent of sending a message
of size $\log_{2}L$ classical bits.

Each party has a number $k_{n}$ (or equivalently $\phi_{n}$). The
$n$th party receives $S$ from the ($n-1$)th party. This will be in a certain
state
$l_{n-1}\in\{1,2,...,L\}$. Since the first party receives no
information we will put $l_{0}=1$ so we can still use this
notation.

The most general protocol the $n$th party can follow consists of
selecting a function $f_{l_{n-1}}^{n}(k_{n})$ which outputs a number
$l_{n}\in\{1,2,...,L\}$. He then prepares system $S$ in this state
$l_{n}$, sending it on to the ($n+1$)th party. The protocols adopted
by all the parties must enable the last party ($N$th) to observe
system $S$ in state $l_{N-1}$, and use this together with his local
information $k_{N}$ to determine whether $m$ is odd or even.

Next we show that for a perfect classical protocol
is is necessary to have $L\geq2N-1$. Since $N$ and $K$ are arbitrary
and can be chosen to be as large as we want, the number of bits
encoded in system $S$ must also be arbitrarily large.

We start by choosing the number of distinguishable states $L$ to be
less than what is necessary to encode each party's number; so let us
set $L=2K-1$. The first party sends message
$l_{1}=f_{0}^{1}(k_{1})$ to the second party. However, this function
cannot be one to one since $L<2K$. Thus, for some $l_{1}$, there must
exist $a \ne b$ such that
$f_{0}^{1}(a)=f_{0}^{1}(b)=l_{1}$. Because of that, when the second
party receives the message $l_{1}$ he does not know whether $k_{1}=a$
or $k_{1}=b$.

The $n$th party will try to enable the next one to learn the value of the
partial sum $(k_{1}+k_{2}+\dots+k_{n})\mbox{ mod }2K$, as this is
the data that the last party needs to know in order to solve the
problem. However, we have just seen that after just one communication step, there are already two different values of the
first party's data $k_1$ assigned to a particular message $l_1$. The
idea of the proof is to show that this uncertainty increases with each
communication step, until there exists a message $l_{N-1}$
that leads the last party to error.

Consider the $n$th party. Let $A_{n}$ be the set of distinct values of
the partial sum $(k_{1}+k_{2}+\dots+k_{n-1})\mbox{ mod }2K$ which are
consistent with the message $l_{n-1}$ he has received. In the case
of the last party, we identify here a possible cause for a flawed
protocol. If $|A_{N-1}|$ has elements that differ by $K$, there will
exist one value of $k_N$ for which the last party will be unable to
solve the problem with certainty.

The $n$th party will send $l_{n}=f_{l_{n-1}}^{n}(k_{n})$ to the
($n+1$)th party. For some $l_{n}$ there must exist distinct $a$ and
$b$ such that $l_{n}=f_{l_{n-1}}^{n}(a)=f_{l_{n-1}}^{n}(b)$ (since
$L<2K$). If the $(n+1)$th party receives this $l_{n}$ then
\begin{equation}
|A_{n+1}|\geq|A_{n}\oplus\{a,b\}|,\label{anplus1}
\end{equation}
where $|A|$ denotes the number of elements of the set $A$ and $A\oplus
B$ is the set of all distinct sums, modulo $2K$, of one element from
$A$ and one element from $B$. Already we see that the number of
elements of $|A_{n}|$ cannot decrease as $n$ increases.

Let us now prove that any successful protocol must have
\begin{equation}
|A_{n}\oplus\{a,b\}|\geq|A_{n}|+1 \hspace{2 mm}, \label{gthana}
\end{equation}
which together with (\ref{anplus1}) means that $|A_{n+1}|$ is strictly
larger than $|A_n|$.

The proof will be by contradiction. Let us thus assume that there is a
successful protocol in which eq. (\ref{gthana}) is false. This is equivalent to
saying that $|A_{n}\oplus\{a,b\}|=|A_{n}|$, as $|A_{n}\oplus\{a,b\}|$ cannot be
less than $|A_{n}|$. Note also that
$|A_{n}|=|A_{n}\oplus\{a\}|=|A_{n}\oplus\{b\}|$. From our assumption and from 
writing
\begin{equation}
A_{n}\oplus\{a,b\}=(A_{n}\oplus\{a\})\cup(A_{n}\oplus\{b\}), \label{aunionb}
\end{equation}
it follows that $A_{n}\oplus\{a\}=A_{n}\oplus\{b\}$ (if they were not
the same, their union would have more elements than each of them separately).

Since we are doing modulo arithmetic we can think of the members of
the set $A_{n}$ as being arranged around a circle. Then the effect of
the $\oplus\{a\}$ operation is to rotate them all forward by $a$ and
similarly for the $b$ case. Hence, $A_{n}\oplus\{a\}=A_{n}\oplus\{b\}$
implies that $A_{n}=A_{n}\oplus\{\left|  b-a\right|  \}$. Put
$|b-a|=\Delta_{1}$. We can apply this shift as many times as we
like. Hence, $A_{n}=A_{n}\oplus \{i\Delta_{1}\}$ where
$i=1,2,\cdots$. It is possible that $\Delta_{1}$ divides $2K$ in which
case it is a period of the set $A_{n}$. If it is not period then we
can still prove that $A_{n}$ must have a period. To see this note that
either $\Delta_{1}$ is a period or there exists an integer $i_{1}$
such that $0< i_{1}\Delta_{1}\mod2K<\Delta_{1}$. Put
$\Delta_{2}=i_{1} \Delta_{1}\mod 2K$. Now,
$A_{n}=A_{n}\oplus\{i\Delta_{2}\}$ and hence either  $\Delta_{2}$ is a
period or there exists an integer $i_{2}$ such that $0<
i_{2}\Delta_{2} \mod2K<\Delta_{2}$. We can continue in this way
generating a sequence of $\Delta_{j}$'s. At some point this sequence
must terminate since $\Delta_{j+1}<\Delta_{j}$ and
$\Delta_{j}\geq1$. The last member of the sequence must be a period of
$A_{n}$. Let this period be $v$.

As we have chosen $K$ to be a power of two, this divisor $v$ must be
either equal to one or to a power of two. These two possibilities
entail the existence of elements of $A_n$ which differ
by $K$, leading to an error. Since we assumed from the beginning that
there was a successful protocol in which eq. (\ref{gthana}) was false,
to avoid a contradiction we are forced to conclude that
eq. (\ref{gthana}) must be true.

Thus, in  any successful protocol eq. (\ref{gthana}) must hold for all
$n<N$. The last party then may receive a message for which
$|A_{N-1}|=N$. If this number is larger than $K$ then there must exist
elements in $A_{N-1}$ separated by $K$, leading to an error. So we need $K \ge 
N$. Now let us remember that we fixed the number of distinguishable states in system $S$ to be $L=2K-1$. This means that any flawless protocol requires that $L \ge2N-1$. As $K$ and $N$ can be chosen at will, this proves that any classical protocol for the task proposed requires a system $S$ with an arbitrarily large number of distinguishable states.

\noindent\textbf{Discussion. }We have proved that any classical system
$S$ with a finite number of distinguishable states fails to provide a
flawless protocol for this task. Unlike classical bits, a qubit has an infinite, continuous set of pure states, i.e. states which are not equivalent to a statistical mixture of two or more distinct states. It is this property that allows for the gap we have proved in communication power between a qubit and any classical system with only a finite number of distinghishable states.

Interestingly, this also turns out to be the distinguishing trait
between quantum theory and classical probability theory. This has been
shown in \cite{Hardy01}, where a set of five axioms for quantum
theory was presented. The first four are consistent both with
classical probability theory and quantum theory. The last axiom
requires exactly that there exist such continuous transformations
between pure states and it is this that rules out classical
probability theory and gives us quantum theory.

Our result places constraints on hidden-variable theories capable of
reproducing the physics of a single qubit. It implies that the hidden
variable must be able to take on an arbitrarily large number of
values. While this is indeed the case for Bell's deterministic
hidden-variable model for a single spin-half particle \cite{Bell66},
here we have shown that this is unavoidable for perfect agreement with
quantum theory.

It is instructive to revisit the classical protocol and spot the
reason why the rod could be replaced by a qubit. The key point to note is that the classical protocol with a rod works just as well if we never acquire information about its position until the final point $B$. There, constraint (\ref{constraint}) guarantees the rod will be in one of two distinguishable states giving the answer. The rod
offers us more than we actually need: it is not necessary that the intermediary states be distinguishable, only pure. This is precisely what a qubit offers us.

The discrete protocol used in our proof can be interpreted as a result
in space complexity theory \cite{AmbainisF98, Watrous99,
TravaglioneNWA02}, which deals with the size of writable computer memory necessary for a computation. We can define a function
$g:n\rightarrow k$, where $n\in\{1,2,...,N\}$ represents each of the
parties and $k\in\{0,1,...,2K-1\}$ represents the numbers each party
has. The constraint on the integral of $\phi(x)$ translates as a
constraint on the function: $\sum_{j=1}^{N}g(j)\mbox{ mod }2K=0$
or $K$. The computational task now is to distinguish between these two
possibilities.

If $K>N$, a change in a single party's number can change the
sum $\mbox{ mod }2K$ from $0$ to $K$ or vice-versa. This means
that $N$ function evaluations are necessary for a flawless
computation, and we have shown that the classical memory (system $S$)
must have at least log$_{2}[2N-1]$ bits. With the same time complexity
(measured in number of function evaluations), a single qubit of
quantum memory can substitute for that large classical memory. This
unbounded difference contrasts with the separation between one qubit
against just two classical bits found in \cite{TravaglioneNWA02},
where there was no limit to the number of function evaluations
allowed.

Our result shows that quantum computers can be more
space-efficient than classical computers, without necessarily
sacrificing time-efficiency. While the results here are for deterministic computation, it would be interesting to compare the accuracy and number of gates needed for an error-bounded computation, in the analog, digital or quantum models of computation. Some preliminary results on this problem were reported in \cite{Galvao02b}.

Yet another interpretation of our result sheds light on the cost of
simulating entanglement with classical communication. Suppose that we
replace each quantum communication step in the discrete version of the
protocol by a teleportation step \cite{BennettBCJPW93}. This would still give us the correct result at the end, but now using only two bits
of one-way classical communication, plus entanglement. Our bound then
implies that no finite amount of one-way classical communication can
perfectly match the communication advantage provided by the entangled
particles. This striking result adds to previous work aiming at
quantifying the communication resources provided by entangled quantum
systems \cite{BrassardCT99, CerfGM00, MassarBCC01}.

The advantage of quantum over classical communication for some
distributed computation problems is amenable to experimental test. In
ref. \cite{Galvao02} an information processing task similar to the
one we analyse here was shown also to be solvable with a single qubit
of sequential communication. It was shown further that the advantage
of one qubit over one classical bit increases with the number of
parties, allowing for an experimental implementation with low quantum detection efficiency.

\textbf{Acknowledgments.} We acknowledge support from the Brazilian
agency Coordena\c{c}\~{a}o de Aperfei\c{c}oamento de Pessoal de Nivel
Superior (CAPES) and from the Royal Society of London.

\end{document}